\documentclass[twocolumn,aps,showpacs,superscriptaddress,floatfix]{revtex4}
\usepackage{amsmath,amssymb,graphicx}
\usepackage{subfigure}
\usepackage{color}
\definecolor{B}{RGB}{1,1,1}
\begin{document}
\title{Granular Gases under Extreme Driving}
\author{W.~Kang}
\email{wfkang@gmail.com}
\affiliation{Department of Physics, University of Massachusetts,
Amherst, Massachusetts 01003 USA}
\author{J.~Machta}
\email{machta@physics.umass.edu}
\affiliation{Department of Physics, University of Massachusetts,
Amherst, Massachusetts 01003 USA}
\author{E.~Ben-Naim}
\email{ebn@lanl.gov}
\affiliation{Theoretical Division and Center for Nonlinear Studies,
Los Alamos National Laboratory, Los Alamos, New Mexico 87545 USA}
\begin{abstract}
We study inelastic gases in two dimensions using event-driven
molecular dynamics simulations. Our focus is the nature of the
stationary state attained by rare injection of large amounts of energy
to balance the dissipation due to collisions. We find that under such
extreme driving, with the injection rate much smaller than the
collision rate, the velocity distribution has a power-law high energy
tail. The numerically measured exponent characterizing this tail is in
excellent agreement with predictions of kinetic theory over a wide
range of system parameters.  {\color{B} We conclude that driving by rare but
powerful energy injection leads to a well-mixed gas and constitutes an
alternative mechanism for agitating granular matter.} In this distinct
nonequilibrium steady-state, energy cascades from large to small
scales. Our simulations also show that when the injection rate is
comparable with the collision rate, the velocity distribution has a
stretched exponential tail.

\end{abstract}


\maketitle

Granular materials are ubiquitous in nature, but nevertheless,
fundamental understanding of the properties of granular materials
presents many
challenges~\cite{jaeger96,kadanoff99,gennes99,ig,bp}. Underlying these
challenges are structural inhomogeneities, macroscopic particle size,
and energy dissipation, all of which are defining features of granular
matter.

Equilibrium gases have Maxwellian velocity distributions. Due to the
irreversible nature of the dissipative collisions, granular gases are
out of equilibrium. Indeed, non-Maxwellian velocity distributions are
observed in a wide range of experiments in driven granular matter
including in particular shaken grains
\cite{olafsen98,gzb,losert99,cafiero2000,rm,blair01,
aranson02,prevost02,wp,fm,tmhs,kohlstedt05}. In such experiments,
energy is injected {\em over a wide range of scales} and the measured
velocity distribution has a stretched exponential form. To a large
extent, a kinetic theory where energy injection through the system
boundary is modeled by a thermostat successfully describes these
nonequilibrium steady-states~\cite{ep,brey03,ve,krb}.

Furthermore, theoretical studies suggest that the steady-state is
controlled primarily by the ratio between the energy injection rate
and the collision rate \cite{mackintosh04,mackintosh05}. When the
injection rate is much larger than the collision rate, the velocity
distribution is Maxwellian.  However, when the injection rate is
smaller than the collision rate, the velocity distribution is
non-Maxwellian, and has a stretched-exponential tail.

In this study, we focus on the limiting case where the injection rate
is vanishingly small and energy is injected at {\em extremely
large velocity scales} \cite{bm,bmm}. Under such extreme driving, the
injected energy cascades down from large velocity scales to small
scales and thereby counters the dissipation by collisions. Kinetic
theory shows that in the stationary state, the velocity distribution
has a power-law high energy tail. These theoretical predictions were
supplemented by Monte-Carlo simulations of the homogeneous Boltzmann
equation where spatial correlations are ignored. However, such
nonequilibrium steady-states have yet to be observed using more
realistic molecular dynamics simulations. In this work we carry out
extensive molecular dynamics simulations to investigate the behavior
of inelastic hard disks under extreme driving.

{\color{B} The goal of this investigation is to establish whether
extreme driving is a feasible mechanism for driving granular matter.}
Our main result is that under rare but powerful injection of energy, a
granular gas indeed reaches a stationary state that is characterized
by a power-law velocity distribution. Moreover, our simulations
quantitatively confirm the predictions of the kinetic theory as the
exponent characterizing the tail of the distribution is validated over
a wide range of parameters.  Our results show that extreme driving is
a feasible mechanism for agitating granular matter.

\noindent{\em Kinetic Theory.} Our starting point is the observation
that the purely-collisional homogeneous Boltzmann equation supports
stationary solutions \cite{bm, bmm,etb,bz}. The evolution equation for the
velocity distribution $f({\bf v})$ of inelastic hard disks takes the
form
\begin{eqnarray}
\label{nonlinear}
\frac{\partial f({\bf v})}{\partial t}\!\! &=&\!\!\iiint d\hat{\bf n}\,d{\bf
u}_1\, d{\bf u}_2 \left|({\bf u}_1-{\bf u}_2)\cdot\hat{\bf n}\right|
f({\bf
u}_1)f({\bf u}_2)\nonumber\\
&\times&[\delta({\bf v}-{\bf v}_1)-\delta({\bf v}-{\bf u}_1)],
\end{eqnarray}
with $\hat{\bf n}$ the impact direction. This Boltzmann equation is
supplemented by the inelastic collision rule which specifies the post-collision velocities ${\bf v}_{1,2}$ as a linear combination of the
pre-collision velocities ${\bf u}_{1,2}$,
\begin{equation}
\label{rule}
{\bf v}_{1,2}={\bf u}_{1,2}-\frac{1+r}{2}({\bf u}_{1,2}-{\bf
u}_{2,1})\cdot\hat{\bf n}\,\hat{\bf n}.
\end{equation}
In an inelastic collision, the normal component of the relative
 velocity reverses sign and is scaled down by the restitution
 coefficient $0\leq r\leq 1$, \hbox{$({\bf v}_1-{\bf v}_2)\cdot
 \hat{\bf n} =-r ({\bf u}_1-{\bf u}_2)\cdot \hat{\bf n}$}. The energy
 loss equals \hbox{$\Delta E=-\frac{1-r^2}{4}|({\bf u}_1-{\bf
 u}_2)\cdot\hat{\bf n}|^2$}.

The collision rule \eqref{rule} simplifies to a ``fragmentation'' rule
\hbox{${\bf u}\to ({\bf w}_1,{\bf w}_2)$} for collisions involving one
extremely energetic particle with velocity ${\bf u}$ and a second,
implicit, particle with speed much less than $|{\bf u}|$.  The
post-collision velocities \hbox{${\bf w}_1=\frac{1+r}{2}{\bf
u}\cdot\hat{\bf n}\,\hat{\bf n}$} and \hbox{${\bf w}_2={\bf
u}-\frac{1+r}{2}{\bf u}\cdot\hat{\bf n}\,\hat{\bf n}$} follow by
substituting ${\bf u}_1=0$ and ${\bf u}_2={\bf u}$, respectively,
into \eqref{rule}.  In the limit $|{\bf v}|\to\infty$, the nonlinear
Boltzmann equation \eqref{nonlinear} becomes linear and its stationary
form is
\begin{eqnarray}
\label{linear}
0\!=\!
\iint d\hat{\bf n}\,d{\bf u} \left|{\bf u}\cdot\hat{\bf n}\right|f({\bf u})
\left[\delta({\bf v}\!-\!{\bf w}_1)\!+
\!\delta({\bf v}\!-\!{\bf w}_2)\!-\!\delta({\bf v}\!-\!{\bf u})\right].
\end{eqnarray}

For arbitrary dimension and for arbitrary collision parameters, this
linear and homogeneous equation admits the power-law solution
\begin{equation}
\label{powerlaw}
f(v)\sim v^{-\sigma}.
\end{equation}
In two-dimensions, the subject of our investigation, the exponent
$\sigma$ obeys the transcendental equation \cite{bm}
\begin{equation}
\label{sigma}
    \frac{1 -
    \!{}_2F_1\left(\frac{3-\sigma}{2},1,\frac{3}{2},1-(\frac{1-r}{2})^2\right)}
    {\left(\frac{1+r}{2}\right)^{\sigma -3}} = \frac{\Gamma(
    \frac{\sigma-1}{2}) \Gamma(
    \frac{3}{2})}{\Gamma(\frac{\sigma}{2})},
\end{equation}
where $_{2}F_{1}$ is the hypergeometric function. The exponent
$\sigma$ grows monotonically with the restitution coefficient. The
limiting values are $\sigma=4.14922$ for completely inelastic
collisions ($r=0$) and $\sigma=5$ in the elastic limit ($r \rightarrow
1$).

The power-law distribution \eqref{powerlaw} is a stationary solution
of the {\em linear} Boltzmann equation \eqref{linear}. Yet, Monte
Carlo methods show that the full {\em nonlinear} Boltzmann equation
\eqref{nonlinear} does admit a stationary solution with a tail given
by \eqref{powerlaw}. These numerical solutions are computed by
injecting energy at a rate that is much smaller than the collision
rate. In an individual injection event a randomly chosen particle is
given a velocity much larger than the typical velocity.  Such extreme
driving maintains a steady-state in which energy injection balances
energy dissipation.

The physical mechanism underlying these driven steady-states is a
cascade in which a high energy particle collides with a particle of
typical energy yielding two high energy particles, each with energies
less than that of the original high energy particle.  These two high
energy particles produce two more high energy particles, again by
collisions with the much more numerous particles with typical
energies, and so on.

\noindent{\em Scaling Analysis.} Interestingly, there is a family of
steady-states generated by extreme driving. If $f(v)$ is a stationary
solution of \eqref{nonlinear}, then $v_0^{-2}f(v/v_0)$ with the
arbitrary typical velocity $v_0$ is also a stationary solution because
the collision rule \eqref{rule} is invariant under the scale
transformation ${\bf v}\to {\bf v}/v_0$. The energy injection rate
$\gamma$, the velocity injection scale $V$, and the typical velocity
$v_0$ are related by the energy balance requirement.

We relate these three quantities by a heuristic argument and first
note that the energy injection rate is simply $\gamma V^2$.  Also, we
anticipate that the velocity distribution is truncated at the
injection scale $V$. The energy dissipation is dominated by the tail
of the distribution and is controlled by the upper cut-off $V$. We
estimate the dissipation rate $\Gamma$ as follows
\cite{norm},
\begin{eqnarray}
\label{dissipation}
\Gamma \sim
\rho\!\int^V\!\!\! v\cdot v^2 \frac{1}{v_0^2} 
f\left(\frac{v}{v_0}\right) \,v\,dv \sim
\rho V^{3}(V/v_0)^{2-\sigma},
\end{eqnarray}
where $\rho$ is the particle density.  In the integrand, the first
term $v$ accounts for the collision rate, and the second term $v^2$
accounts for the energy dissipation in an inelastic collision.  The
integration is performed using the velocity distribution
\eqref{powerlaw}, and since $\sigma<5$ the dissipation is indeed
dominated by the high velocity tail of the distribution.  Balancing
energy injection with dissipation we obtain a relationship between the
injection rate $\gamma$, the injection scale $V$, and the typical
velocity $v_0$,
\begin{equation}
\label{injection}
\gamma\sim \rho\, V (V/v_0)^{2-\sigma}.
\end{equation}

Since the collision rate is proportional to $\rho\,v_0$, the
dimensionless ratio $\psi$ of the injection rate to the collision rate
scales as a power of the velocity ratio $V/v_0$,
\begin{equation}
\label{ratio}
\psi \sim (V/v_0)^{3-\sigma}.
\end{equation}
Since $\sigma>3$, we expect a wide power-law range, $V\gg v_0$, when
the injection rate is much smaller than the collision rate, $\psi\ll
1$. {\color{B} There is no lower cutoff on the injection rate $\psi$,
below which power-law is not observed; the smaller is $\psi$, the
broader the power-law range.} When $\psi$ is order one, the velocity
distribution no longer has a power-law tail, and of course, when
$\psi\gg 1$, the velocity distribution should simply mirror the
distribution of injected velocities.

\noindent{\em Molecular Dynamics Simulations.} We used molecular
dynamics~\cite{alder59} to simulate inelastic hard disks in a square
box with elastic walls. In these event driven simulations, upon
impact, the velocities of the colliding particles are updated
according to the collision rule \eqref{rule}. Subsequent to each
collision, we identify the time and location of the next collision.
The particles undergo purely ballistic motion between two successive
collisions.

We implemented the following velocity-dependent restitution
coefficient \cite{goldman98}
\begin{equation}
\label{rc}
    r(\delta_n) =
\begin{cases}
    1 - (1 - r) (\delta_n/v_c)^{3/4} & \text{$\delta_n < v_c$},  \\
    r & \text{$\delta_n \geq v_c$},
\end{cases}
\end{equation}
where $\delta_n = (\mathbf{v_1} - \mathbf{v_2}) \cdot
\mathbf{\hat{n}}$ is the normal component of the relative
velocity. Here, $r$ is the nominal value of the restitution
coefficient, valid at large velocities, and $v_c$ is the cutoff
velocity, below which collisions become elastic.  With this realistic
restitution coefficient \cite{luding96,bizon98}, we avoid inelastic
collapse where an infinite number of collisions can occur in a finite
time \cite{my}.  Typically, we set $v_c$ much smaller than the typical
velocity $v_0$, but for small restitution coefficients, we must set
$v_c$ comparable to $v_0$ to avoid inelastic collapse.

To maintain a steady-state, we periodically boost a {\em single}
randomly-selected particle to a large, random velocity. These
injection events are rare and they are governed by a Poisson process
with rate $\gamma$, that is, with probability $\gamma\,dt$ injection
is implemented during the time interval $[t,t+dt]$. The injection
speed is selected from a Gaussian distribution with zero mean and
standard deviation $V$. By taking a long-time average, we confirmed
that the total energy approaches a constant, and hence, that the
system reaches a statistical steady-state where energy injection and
energy dissipation balance. Moreover, the velocity distributions were
produced by sampling particle velocities at a very large number
($10^8$) of equally spaced time intervals.  {\color{B} We stress that
the velocities are sampled at time intervals that are completely
uncorrelated with either collision events or injection events. We
tested that our sampling produces robust velocity distributions, and
that the velocity distribution, representing an average over the
entire system, is truly stationary. In particular, the system does not
enter the homogeneous cooling state \cite{pkh} in between the rare
injection events.}

\begin{figure}[t]
\includegraphics[width=0.4\textwidth]{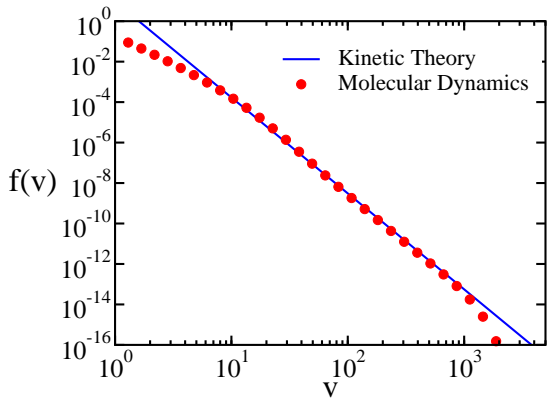}
\caption{The velocity distribution $f(v)$ versus the velocity
$v$. Molecular dynamics simulation results (bullets) are compared with
the power-law tail predicted by kinetic theory (solid line).}
\label{fig-fv}
\end{figure}

We performed numerical simulations using a system of $N=10^3$
identical particles with diameter $2R=1$ in a square box of size
$L=400$, corresponding to the low area fraction
$\phi=N\pi\,(R/L)^2=4.9 \times 10^{-3}$. Unless noted otherwise, these
parameters are used throughout this study.  Energy was injected at
rate $\gamma= 5 \times 10^{-7}$ and the injection scale was $V=850$.
First, we considered weakly-inelastic particles, $r=0.9$, with the
cutoff $v_c=0.1$. With these parameters, the injection rate is much
smaller than the measured collision rate and their ratio is
\hbox{$\psi=1.5 \times 10^{-5}$}. Consequently, there is substantial
scale separation between the typical velocity $v_0$ and the injection
velocity $V$. Over this range, the steady-state velocity distribution
obtained by the molecular dynamics simulations has a power-law
high energy tail as in \eqref{powerlaw} and the exponent $\sigma=4.74$
is in very good agreement with the kinetic theory prediction given by
\eqref{sigma}, $\sigma=4.74104$ (see figure \ref{fig-fv}). We also
confirmed that the ratio $V/v_0\approx 10^2$ is consistent with the
scaling estimate \eqref{ratio}.

\begin{figure}[t]
\includegraphics[width=0.4\textwidth]{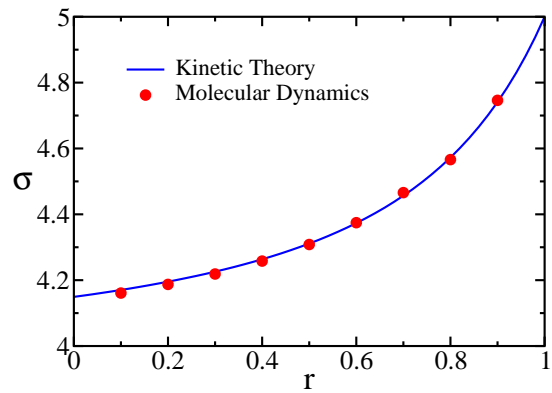}
\caption{The exponent $\sigma$ versus the restitution coefficient
$r$. The molecular dynamics results (bullets) are compared with the
kinetic theory predictions (solid line).}
\label{fig-sigma}
\end{figure}

Next, we varied the restitution coefficient and repeated the
simulations.  Over the entire parameter range \hbox{$0.1\leq r \leq
0.9$}, we find stationary velocity distributions with a power-law
tail. In general, the ratio $V/v_0$ is consistent with the scaling
relation \eqref{ratio}. Moreover, the exponent $\sigma$ obtained from
the molecular dynamics simulations is in excellent agreement with the
kinetic theory predictions \eqref{sigma} for all restitution
coefficients (figure \ref{fig-sigma}). We thus arrive at our main
result that at least for dilute gases, extreme driving in the form of
rare but powerful energy injection generates a steady-state with a
broad distribution of velocities. The tail of the velocity
distribution is power-law and the characteristic exponent is
nonuniversal as it depends on the restitution coefficient.

The fact that kinetic theory holds shows that, to good approximation,
the gas is well-mixed.  We comment that it is remarkable that extreme
driving results in a well-mixed gas. On short time scales, energy
injection clearly generates spatial correlations because a
just-energized particle transfers much of its energy to nearby
particles by inelastic collisions. Yet, on larger time scales
energetic particles break coherent structures which are known to be
the consequence of inelastic collisions \cite{gz}. While these two
mechanisms have opposite effects, the simulations indicate that when a
long time average is taken, the latter effect dominates. Thus, energy
injected at extreme velocity scales in a tightly localized region of
space, ends up evenly distributed throughout the system.

{\color{B} Snapshots of the time evolution of the system following energy
injection demonstrate how the inelastic cascade works (figure
\ref{fig:hot}). In the initial stages of the cascade, the injection
affects only a small region in space and, moreover, there are strong
spatial correlations between the velocities of the particles (figure
\ref{fig:hot} a-c).  However, after many inelastic collisions, the
injected energy ends up evenly distributed throughout the system
(figure \ref{fig:hot} d). When a long time average is taken over many
energy injection events at different locations, the system is
maintained in a homogeneous, well-mixed state. Spatial correlations
induced by inelasticity and the injection mechanism do not affect the
predicted power-law velocity distributions.

\begin{figure}[t]
\begin{center}
\subfigure[ ]{
\includegraphics[scale=0.30]{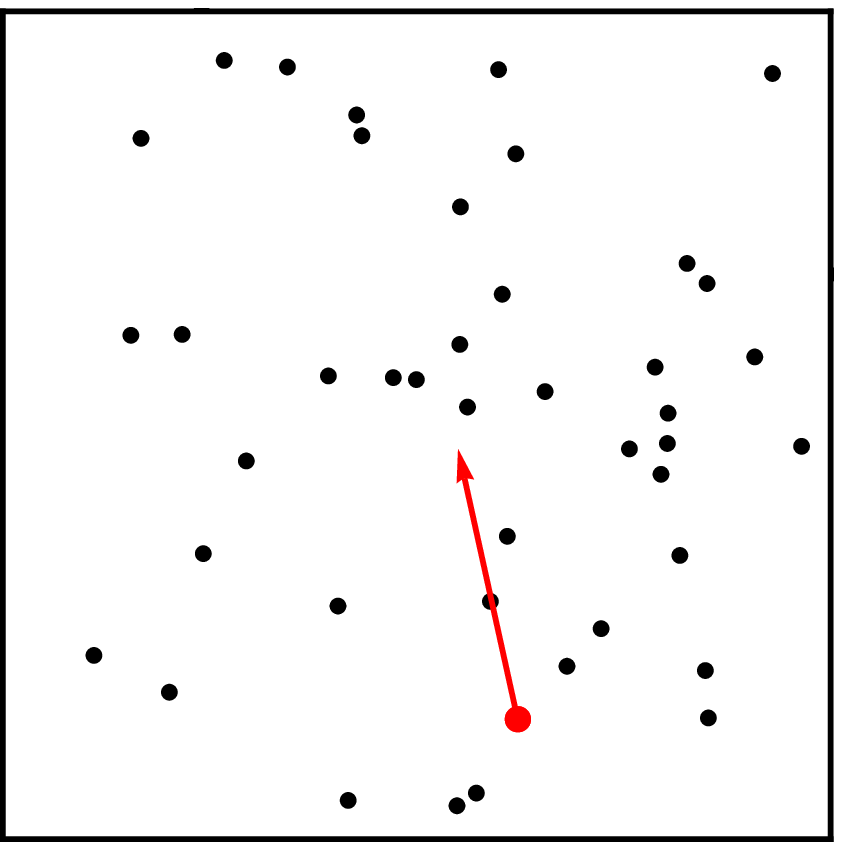}}
\subfigure[ ]{
\includegraphics[scale= 0.30]{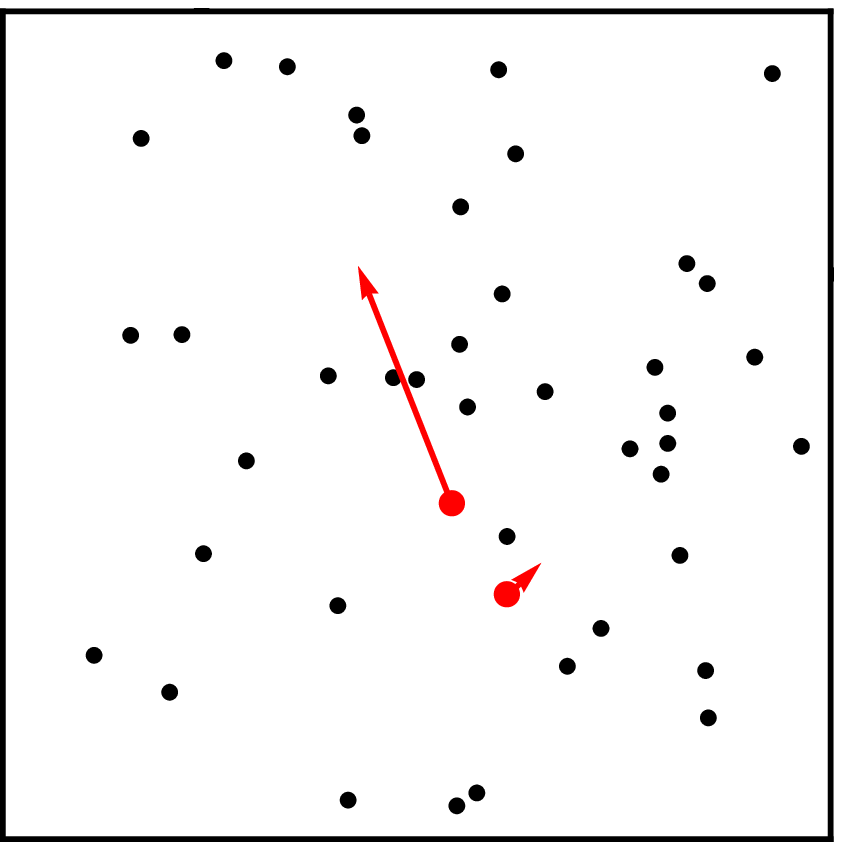}}
\subfigure[ ]{
\includegraphics[scale= 0.30]{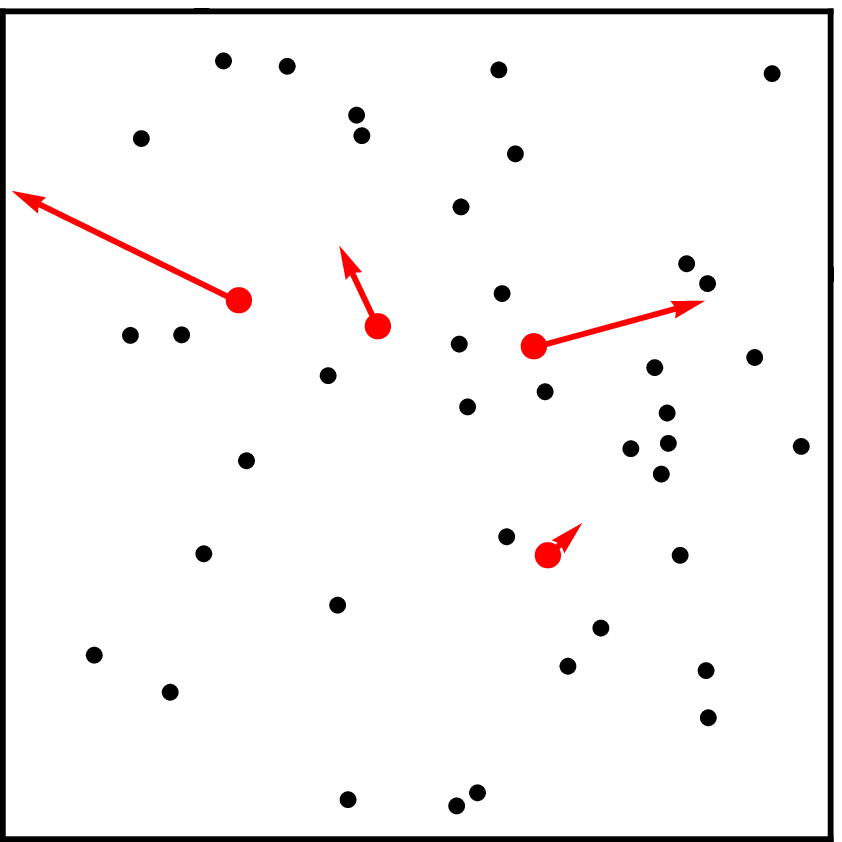}}
\subfigure[ ]{
\includegraphics[scale= 0.90]{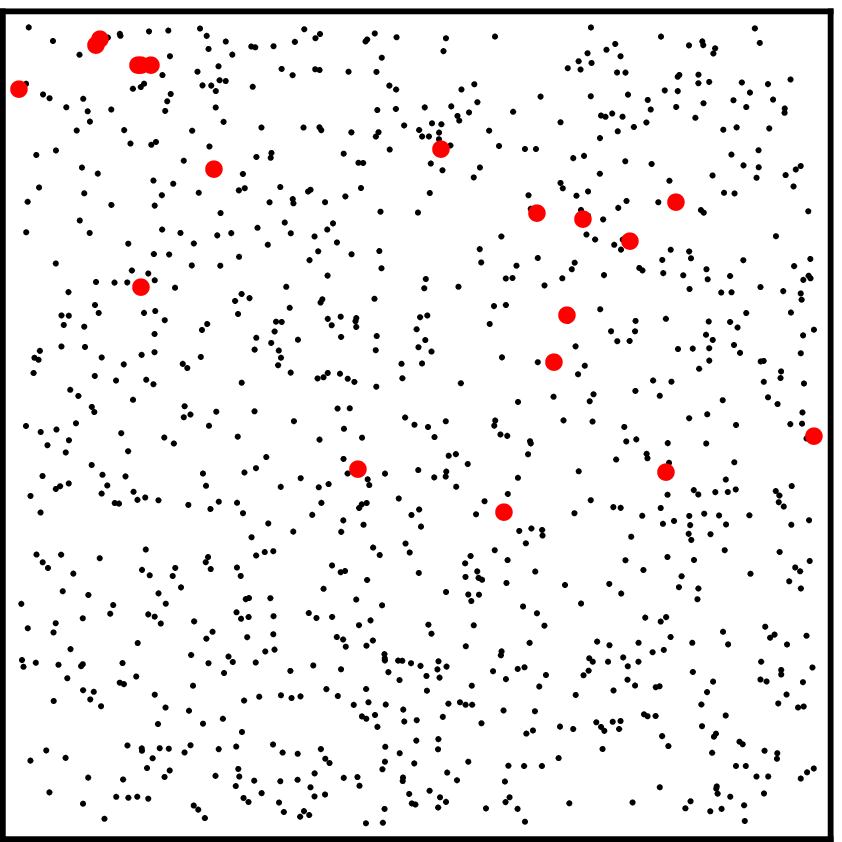}}
\caption{\label{fig:hot} {\color{B} The inelastic energy
cascade. Shown are four time-ordered snapshots of the gas shortly
after an injection event.  The top three figures show a small window
around the injection event in the early stages of the cascade: (a) the
initial energetic particle (red online) (b) two energetic particles
after one collision, (c) four energetic particles after three
collisions. Figure (d) shows the entire system, with energetic
particles shown larger (red online), in a late stage of the cascade.
After many collisions, the injected energy is evenly distributed
throughout the system.  The simulation parameters are $r=0.8$, $N=
1000$, $L=400$, $V=707$ and $\gamma= 2 \times 10^{-6}$.}}
\end{center}
\end{figure}

Therefore, rare, powerful, and spatially localized energy injection is
a unique mechanism of agitating granular gases. This mechanism induces
an extended energy cascade which distributes the injected energy to
the rest of the system. This physical mechanism is different than
energy injection by walls \cite{gzb,losert99} or by an effective
thermostat \cite{ve} or by multiplicative driving \cite{cafiero2000}
where the injected energy directly affects only a small region of
space. In this sense, the energy cascade represents a novel mechanism
for agitating granular matter.

Figure \ref{fig:hot} also illustrates that the velocity distribution
is correlated with injection times for our system so that the
predicted power-law distribution arises only after averaging over many
measurements taken at times that are uncorrelated with the injection
times. However, for very large systems driven with a fixed but small
injection rate per particle there would be many temporally overlapping
but spatially well-separated injection events.  At any instant of time
in a very large system cascades at all stages of development would be
present somewhere in the system and the power law tail would be
time-independent.}

\begin{figure}[t]
\includegraphics[width=0.4\textwidth]{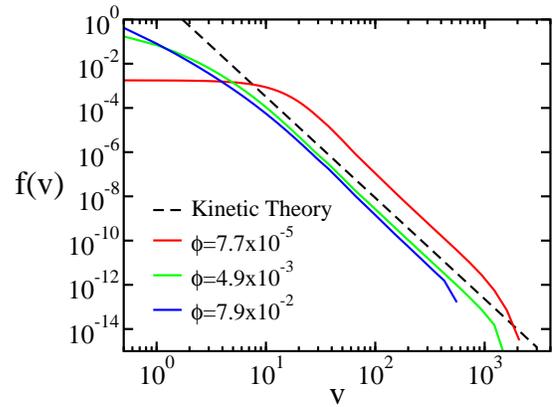}
\caption{The velocity distribution at three different
densities (solid lines). The simulation parameters are: $r=0.8$,
$\gamma=2\times 10^{-6}$, and $V=707$. Eq.~\eqref{powerlaw} with
$\sigma=4.57246$ is also shown as a reference (dashed line). A best
fit to a power-law yields $\sigma=4.6$, $4.6$, and $4.7$ for
$\phi=7.7\times 10^{-5}$, $4.9\times 10^{-3}$, and $7.9\times
10^{-2}$, respectively.}
\label{fig-density}
\end{figure}

We performed additional simulations to test whether the results are
robust with respect to change of parameters. In particular, we varied
the area fraction by fixing the number of particles and varying the
system size. The results shown in figure \ref{fig-density} are for
three different area fractions: $\phi=7.7\times 10^{-5}$,
$\phi=4.9\times 10^{-3}$, and $\phi=7.9\times 10^{-2}$.  {\color{B} The
corresponding values of $\psi$ are $4.7 \times10 ^{-4}$, $5.4
\times10^{-5}$, and $3.4\times10^{-6}$, respectively.  Note that in
all cases $\psi \ll 1$.}  We find the same power-law tail in all three
cases, and the exponent is in good quantitative agreement with the
kinetic theory prediction (figure \ref{fig-density}).   {\color{B} Thus, the
energy cascade mechanism can be realized even at area fractions as
high as $\phi\approx 10^{-1}$ and with $\psi$ as small as order $10^{-6}$.}

\begin{figure}[t]
\includegraphics[width=0.4\textwidth]{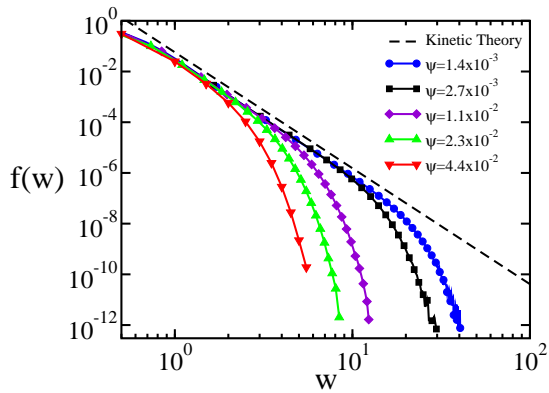}
\caption{The velocity distribution $f(w)$ versus the
normalized velocity variable $w=v/v_0$ with different ratios of
injection rate to collision rate: $\psi=1.4\times 10^{-3}$ (bullets),
$2.7\times 10^{-3}$ (squares), $1.1\times 10^{-2}$ (diamonds),
$2.3\times 10^{-2}$ (up-triangles), and $4.4\times 10^{-2}$
(down-triangles). The restitution coefficient is $r=0.8$. Also shown
is the reference theoretical curve \eqref{powerlaw} with
$\sigma=4.57246$ (dashed line).}
\label{fig-injection}
\end{figure}

\begin{figure}[t]
\vspace{.1in}
\includegraphics[width=0.4\textwidth]{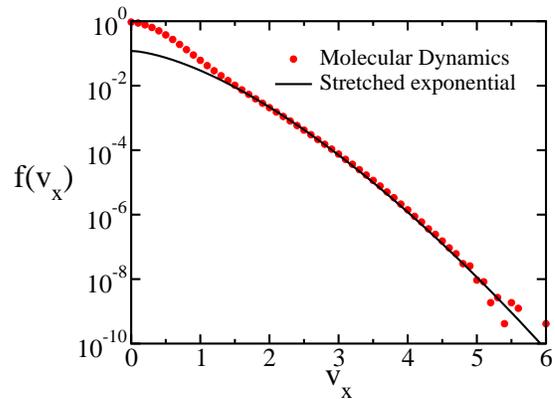}
\caption{The distribution $f(v_x)$ of the horizontal component of the
velocity $v_x$ at a moderate $\psi=5.8\times 10^{-2}$.  These
simulations are performed with $r=0.8$, $\gamma=6\times 10^{-4}$, and
$V=1.41$. The solid line is a best-fit to the stretched exponential
\hbox{$f(v_x)\sim \exp\big(-{\rm const.}\times |v_x|^\zeta\big)$} with
$\zeta=1.52$.}
\label{fig-stretched}
\end{figure}

We also studied the dependence on the ratio $\psi$ between the
injection rate and the collision rate by varying the injection rate
$\gamma$ and the injection velocity $V$. In accord with \eqref{ratio},
we find that the range $[v_0,V]$ of power-law behavior shrinks as
$\psi$ increases. As long as $\psi$ is sufficiently small, the
distribution has a power-law tail (figure \ref{fig-injection}). When
this ratio becomes sufficiently large, the tail is no longer algebraic
and a sharper decay occurs.  For $\psi=5.8\times 10^{-2}$, we find the
stretched exponential tail $f(v_x)\sim \exp\big(-{\rm const.}\times
|v_x|^\zeta\big)$ with $\zeta=1.52$ (figure \ref{fig-stretched}).
This value is consistent with the theoretical value $\zeta=3/2$ for
inelastic gases driven by white noise \cite{ve,krb} and the
experimental value observed in vigorously shaken beads \cite{rm}.
{\color{B} Indeed, in this intermediate injection rate regime, the
energy cascade becomes localized, and frequent, small injections are
similar to white noise driving. On the other hand, these stretched
exponential tails do not relate to those observed in \cite{bmm} as the
system cools down after injection is turned-off.}  We find stretched
exponential tails for the range $10^{-1}\lessapprox \psi\lessapprox
1$.  When the injection rate exceeds the collision rate ($\psi\gg 1$)
the entire distribution becomes Maxwellian as the velocity
distribution simply mirrors the distribution of injected velocities.

Finally, we mention that we even varied the energy injection mechanism
itself. In particular, to implement injection strictly at large energy
scales, we used an ``energy loss counter" to keep track of the total
energy dissipated by collisions since the last injection. When the
dissipated energy equals a fixed large value, we inject this amount of
energy into a single randomly chosen particle \cite{bm,bmm}.  Using
this variant of extreme driving to maintain the steady-state, we found
similar power-law velocity distributions.  {\color{B} From these
studies, we conclude that the parameter $\psi$ controls the velocity
distribution, and that different energy injection mechanisms lead to a
power-law distribution with the very same exponent $\sigma$, as long
as $\psi\ll 1$.}

\noindent{\em Conclusions.} Extensive event-driven simulations show
that under extreme driving in the form of rare but powerful energy
injections, an inelastic gas reaches a steady-state with a broad
distribution of energies. Such driven steady-states are observed for a
wide range of collision parameters, densities, and energy injection
rates, as long as the injection rate is much smaller than the
collision rate.  The velocity distributions have a power-law tail and
the characteristic exponent is in good agreement with the kinetic
theory predictions. When the ratio between the energy injection rate
and the collision rate becomes sufficiently large, the velocity
distribution has a much sharper stretched exponential tail.

We conclude that extreme driving where energy is injected only at very
large scales presents an alternative mechanism for agitating granular
matter and that such driving leads to a fundamentally different
steady-state compared with traditional driving where energy is
injected over all scales. Realizing this driving in experiments is a
challenge because the agitation must be applied only at very large
velocities. One possible mechanism is shooting very fast particles
into the system. While such a system would involve a growing number of
particles, the injection of energetic particles should lead to
transfer of energy from large scales to small scales by a cascade of
collisions.

\begin{figure}[h!]
\begin{center}
\subfigure{
\includegraphics[scale= 0.3]{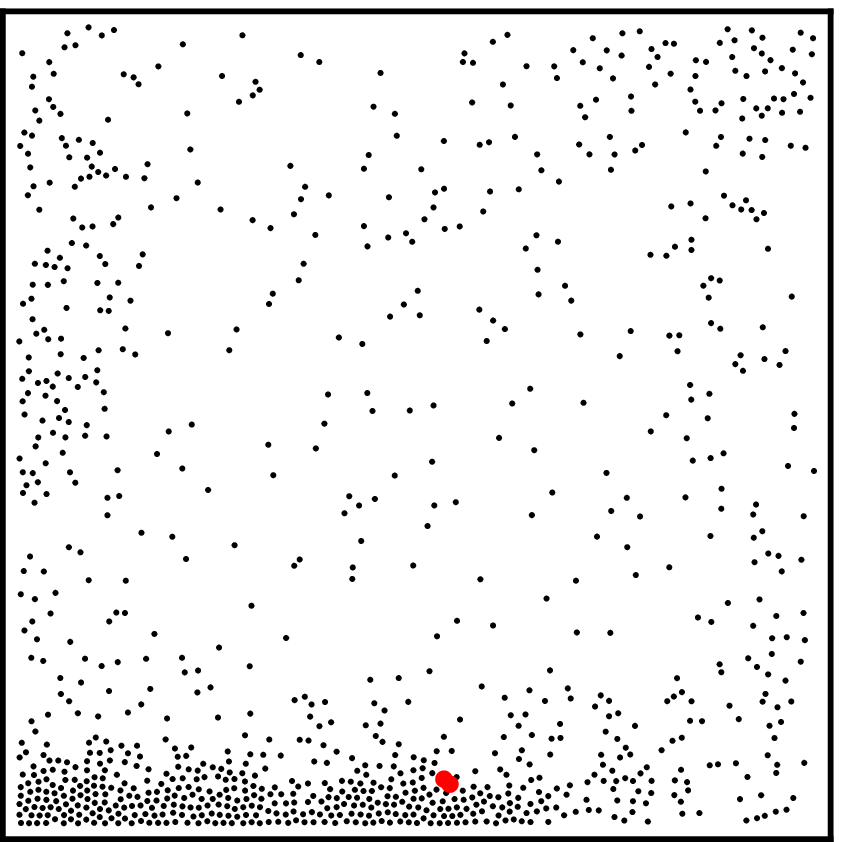}}
\subfigure{
\includegraphics[scale= 0.3]{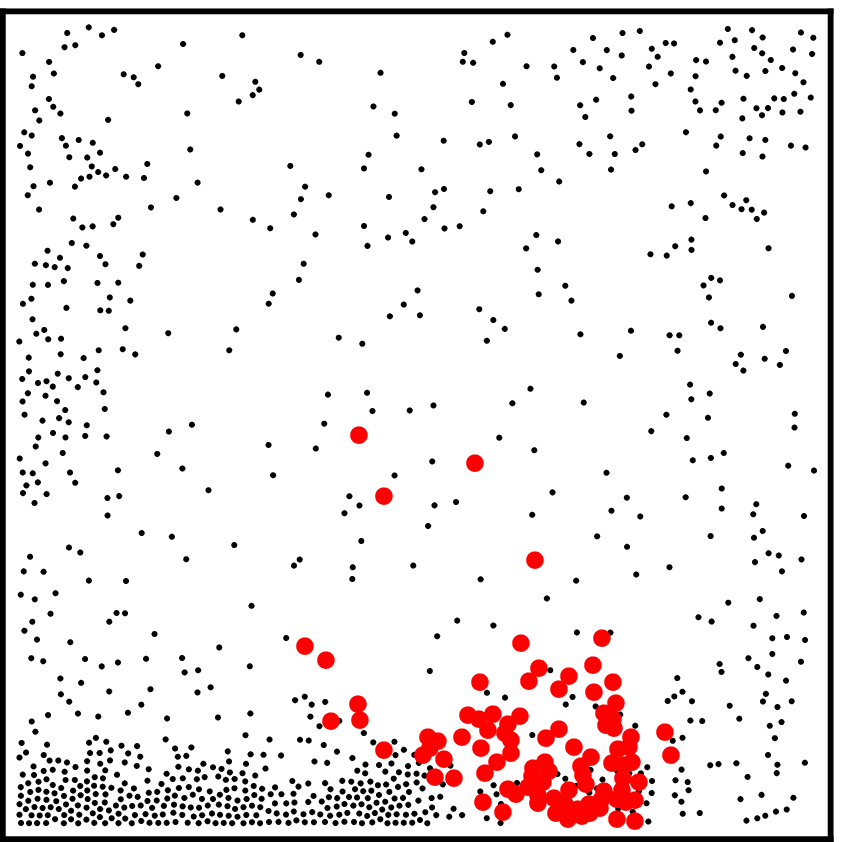}}
\subfigure{
\includegraphics[scale= 0.3]{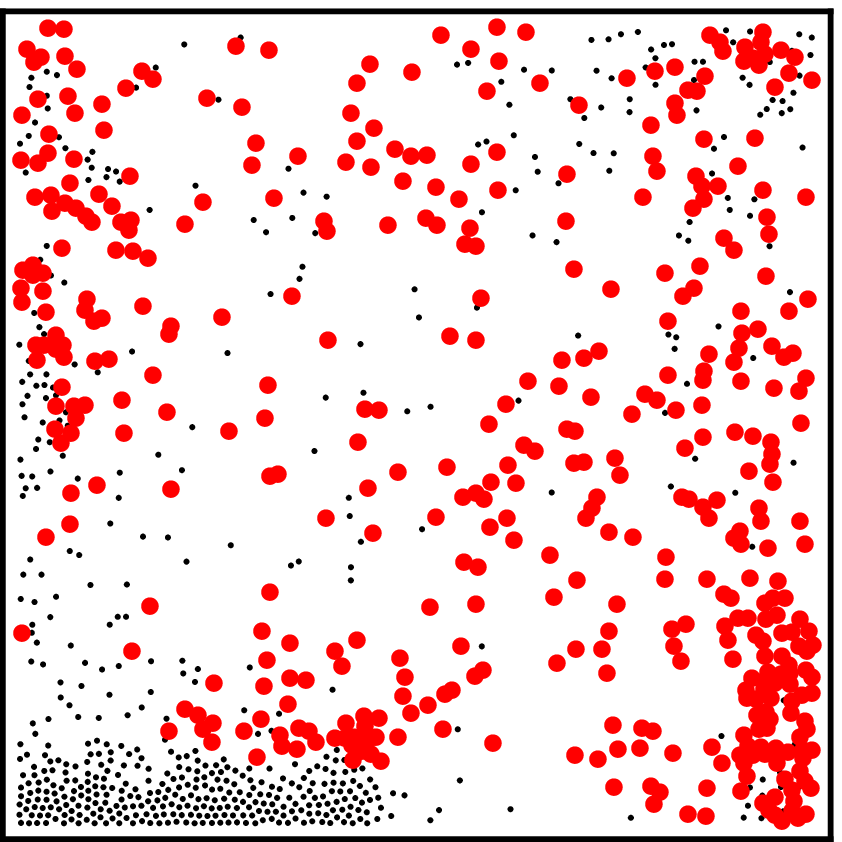}}
\caption{\label{fig:dense} {\color{B} The time development, left to
right, of a cascade in a denser system ($\phi =7.9\times 10^{-2}$ and
the same parameters as in Fig.\ \ref{fig-density}.) Particles with
speeds greater than a fixed threshold are shown larger (red online).}}
\end{center}
\end{figure}

{\color{B} Interestingly, the power-law velocity distribution appears
to hold in systems that include dense clusters. We have observed that
when the density is increased, there is a tendency for clustering near
the walls with no measurable deviation from the predicted power law
(figure \ref{fig:dense}).  In this case injection leads to explosive
breakup of dense regions. It is intriguing that the power-law tail is
quite robust, and extends to situations where the assumptions
underlying the kinetic theory approach can no longer be justified.}

In our simulations, inelastic collapse does not play a role because
collisions become elastic at small relative velocities. Yet, if the
collisions are purely inelastic, there should be a competition between
the formation of high-density regions by inelastic collisions and the
destruction of such clusters by high energy particles. Elucidating
this competition is another possibility for further investigation.
Nonetheless, our simulations suggest that extreme driving generates
well-mixed steady-states despite the fact that this driving is very
inhomogeneous in both space and time.

\noindent{\em Acknowledgments.} We thank Narayanan Menon, Felix Werner
for useful discussions and Hong-Qiang Wang for sharing the molecular
dynamics code. We acknowledge Financial support from NSF grant
DMR-0907235 (JM and WK) and DOE grant DE-AC52-06NA25396 (EB).

\end{document}